\documentclass[draft,11pt]{article}
\usepackage{amssymb,amsthm,amstex}

\begin{document}


\title{Adelic Formulas for Gamma and Beta Functions of~One-Class Quadratic 
Fields:\\  Applications to~4-Particle Scattering\\ String Amplitudes}

\author{V.~S.~Vladimirov\footnote{Steklov Institute of Mathematics, Russian 
Academy of Sciences, ul.~Gubkina 8, Moscow, 117966 Russia.}}

\date{}
\maketitle
\def\({\left(}
\def\){\right)}
\def\[{\left[}
\def\]{\right]}
\def\<{\left<}
\def\>{\right>}
\let\Bbb\mathbb
\def\R{\Bbb R}
\let\q\quad
\let\qq\qquad
\newtheorem*{prp*}{Proposition}
\newtheorem{teo}{Theorem}
\theoremstyle{definition}
\newtheorem{rem}{Remark}
\def\C{\Bbb C}
\def\B{{\mathrm B}}
\let\bar\overline
\let\kappa\varkappa
\makeatletter
\def\case#1{\left\{\,\vcenter{\def\\{\cr}\baselineskip=\normalbaselineskip
        \lineskiplimit=7pt \lineskip=7pt plus 1pt
        \m@th\ialign{$\displaystyle ##\hfil$&\quad##&$\displaystyle ##$
        \hfil\crcr#1\crcr}}\right.}
\def\Mathop#1{\expandafter\def\csname#1\endcsname{\mathop{\mathrm{#1}}\nolimits}}
\@addtoreset{equation}{section}
\renewcommand{\theequation}{\arabic{section}.\arabic{equation}}
\def\a{\alpha} \def\b{\beta}  \def\g{\gamma}  \def\G{\Gamma}  
\def\t{\theta} \def\s{\sigma} \def\u{\tau} \def\T{\Theta}
\Mathop{mod} \Mathop{reg} \Mathop{Re}
\def\RR{\R}  
\def\CC{\C}  
\def\QQ{\Bbb Q}  
\def\KK{\Bbb K}  
\def\9#1{\phantom{#1}}
\def\sgn{\mathop{\mathrm{sgn}}\nolimits}   
\Mathop{Sp} \Mathop{Arg}
\def\sek#1{\par \section{\uppercase{#1}}\par}
\let\nn\nonumber

\abstract{Regularized adelic formulas for gamma and beta functions for 
arbitrary  quasicharacters (either ramified or not) and in an arbitrary  
field of algebraic numbers are concretized as applied to one-class 
quadratic fields (and to the field of rational numbers). Applications to 
4-tachyon tree string amplitudes, to the Veneziano (open strings) and 
Virasoro (closed strings) amplitudes as well as to massless 4-particle 
amplitudes of the Ramond--Neveu--Schwarz superstring and a heterotic 
string are discussed. Certain relations between different superstring 
amplitudes are established.}

\sek{INTRODUCTION}

Here, it is relevant to note that the ideas and methods  developed by the 
great Russian mathematician and physicist Nikolai Nikolaevich Bogolyubov 
(see his selected works [1, 2]) are widely used in the construction of 
$p$-adic mathematical physics. The present work is devoted to the 
regularization of divergent adelic products for string and superstring 
4-particle amplitudes and in many aspects follows the Bogolyubov ideas. In 
fact, the theory of strings and superstrings has evolved from the theory of 
dispersion relations in quantum field theory to whose development a 
fundamental contribution was made by Bogolyubov.

In the last decade, an interest in $p$-adic numbers has considerably 
increased.  These numbers were discovered by the German mathematician 
K.~Hensel [3] about 100 years ago. Until recently, they have not found any 
significant applications although constituted an essential part of number 
theory and theory of representations [4--8]. Unexpected applications of 
these numbers have been associated with the discovery of a non-Archimedean 
structure of space--time at supersmall, the so-called Planck, distances (of 
order~$10^{-33}$\,cm) in quantum gravitation and theory of strings. 
Therefore, one cannot use real numbers with Archimedean structure as 
space--time coordinates at such small distances; one needs other, 
non-Archimedean number fields. An attempt to preserve rational numbers as 
physical observables in appropriate non-Archimedean fields leads us to the 
unambiguous conclusion that the fields of $p$-adic numbers $\QQ_p,$ where 
$p=2,3,5,\ldots$ are prime numbers, can be chosen as such number fields 
(the Ostrovskii theorem!)~[4]. Thus, we arrive at the problem of 
constructing $p$-adic mathematical physics. By now, a considerable progress 
has been made in this direction (see [9, 10], where the motivation, the 
history of the problem, and extensive bibliography were presented).

The construction  of adelic formulas relating $N$-particle scattering 
amplitudes to appropriate $p$-adic amplitudes constitutes a separate field 
of $p$-adic mathematical physics. The first such formulas (without 
regularization) were proposed by Freund and Witten [12] and, independently, 
by Volovich [13] in 1987 as applied to 4-tachyon string amplitudes. 
Mathematically, the problem 
is reduced to the construction of adelic formulas for 
local gamma and beta functions of algebraic number fields. In [14, 15], 
such regularized adelic formulas were constructed in arbitrary fields of 
algebraic numbers for principal (unramified) quasicharacters and in [16, 
17], for ramified ones. In the latter case, these formulas are valid under 
the condition that the quasicharacters of the appropriate group of ideles 
are trivial (i.e., are equal to~1) on the multiplicative group of the 
original field of algebraic numbers [7, 17]. This condition yields explicit 
relations between the parameters of local characters, similar to the 
formulas for a field of rational numbers~$\QQ$~[8]. These relations for 
one-class quadratic fields were obtained in my work~[18] (see Section~2). 
In Section~3, we present regularized adelic formulas for gamma and beta 
functions as applied to one-class quadratic fields (and to the field of 
rational numbers).  As compared with [16, 17], here we remove the technical 
requirements that the ranks of local characters should be identical at all 
finite points (see Remark~5). In Section~4, we apply the formulas obtained 
to 4-tachyon tree string amplitudes: to the Veneziano (open strings) and 
Virasoro (closed strings) amplitudes as well as to the massless 4-particle 
scattering amplitude of the Ramond--Neveu--Schwarz superstring and massless 
scattering amplitudes of four charged particles of a heterotic string. We 
establish certain relations between different superstring amplitudes. We 
also discuss the problem of other possible applications of adelic formulas 
for ramified quasicharacters.

\sek{IDELES AND THEIR QUASICHARACTERS FOR\\ ONE-CLASS QUADRATIC~FIELDS}

First, recall the necessary information from the theory of adeles  (and 
ideles) as applied to the quadratic fields $\QQ(\sqrt d)$, where 
$d\in Z$, $d\neq 0,1$,
is squares free [7,~8, 18,~19]. For this purpose, we first introduce the 
following notation.

Denote by $P_d$ and $S_d$ the sets of prime numbers $p$ such that 
$d\in\QQ_p^{\times 2}$ or $d\not\in\QQ_p^{\times 2}$, respectively.

Let $A_d$ be the adele ring and $A_d^\times$ be the idele group of the 
field $\QQ(\sqrt d)$. An arbitrary adele $X\in A_d$ is expressed as
$$ X=(X_\infty, X_2,X_3,\ldots,X_p,\ldots ),$$ 
where the symbol $X_p$ determines the following points of the ring $A_d$:
\begin{equation*}
\begin{aligned}
X_\infty &=\case{z\in\CC, &&d<0, \\ (x,x')\in\RR^2, &&d>0;} 
\\
X_p &=\case{ (x_p,{x}'_p)\in\QQ_p^2, &&p\in P_d, \\ x_p+\sqrt dx'_p 
\in\QQ_p(\sqrt d), &&p\in S_d;} 
\end{aligned}
\end{equation*}
in this case, for all $p$ starting from a certain value, 
$(x_p,x'_p)\in Z_p^2$, $p\in P_d$, and $X_p\in Z_p(\sqrt d)$, $p\in S_d$.
Here, $Z$,~$Z_p,$ and~$Z_p( \sqrt d)$
are the rings of integers of the fields $\QQ$,~$\QQ_p$, and~$\QQ_p(\sqrt 
d)$, respectively.  The group of ideles $A_d^\times$ is defined similarly. 
Ideles are invertible adeles.

An arbitrary (multiplicative) quasicharacter $\T$ of the group $A_d^\times$ 
on the idele~$X$ is represented as [7, 11, 18]
\begin{equation}
\T (X;\a)=\T_\infty (X_\infty )\prod_{p\in P_d}\t_p(x_p)\t'_p(x'_p)|x_p |_p^{i\a_p}|{x}'_p |_p^{i\a'_p}
\prod_{p\in S_d}\t_p(X_p)|X_p\bar{X}_p |_p^{i\a_p}|X|^\a, 
\qquad \a\in\CC, 
\end{equation}
where \ $\bar X_p=x_p-\sqrt dx'_p$, \ $X_p\bar X_p=x_p^2-d{x'_p}^2$, \ and
\begin{equation}
\T_\infty (X_\infty )=\case{ z^\nu (z\bar z)^{-\nu/2}, &&\nu\in Z,\ d<0,\\ 
\sgn^\nu x\cdot \sgn^{\nu'} x'|x|^{ia}, &&\nu, \nu'\in F_2,\ a\in\RR,\ d>0.}
\end{equation}
Here, $|X|$ denotes the normalization of the idele $X$ and $F_2$ is a 
field of integers modulo~2.

Suppose that the local characters $\t_p$ and ${\t}'_p$ in (2.1) are 
normalized by the conditions
\begin{equation}
\t_p(p)=1, \quad \t'_p(p)=1,\q p\in P_d; \qq
\t_p(q)=1, \q p\in S_d, 
\end{equation}
where $q=q_p$ is the module over the field $\QQ_p(\sqrt d)$, $p\in S_d$.

In (2.1), only a finite number of characters $\t_p$ and $\t'_p$ are 
nontrivial (ramified). Denote the ranks of these characters by 
$\rho_p=\rho(\t_p)$ and $\rho'_p=\rho(\t'_p)$, 
respectively. Denote by~$F$ the set of finite points for which the 
rank of the local character is equal to~0 (unramified points); the set of 
other finite points is denoted by~$R$ (ramified points).

The triviality of the quasicharacter $\T$ on the principal idele $X=I_x$
generated by the number $x\in\QQ^\times (\sqrt d)$ is expressed as
\begin{equation}
\T (I_x;\a)\equiv\T (x)=1, \qq x\in\QQ^\times (\sqrt d). 
\end{equation}
Condition (2.4) implies that, actually, the quasicharacter $\T$ is defined 
over the factor group $A_d^\times/
\QQ^\times (\sqrt d)$, the idele class group over the field 
$\QQ (\sqrt d)$~[7].

Denote the norm of the leading ideal $J$ of the quasicharacter $\T$ by 
$N$ [7]:
\begin{equation}
N=N(J)=\prod_{p\in P_d\cap R}p^{\rho_p+\rho'_p}
\prod_{p\in S_d\cap R}q_p^{\rho_p}. 
\end{equation}

A standard additive character $\chi$ of the adele ring $A_d$ is given by
\begin{equation}
\chi (X)=\chi_\infty (X_\infty )\prod_{p\in P_d}\chi_p(x_p+x'_p)
\prod_{p\in S_p}\chi_p(X_p+\bar X_p), \qquad X\in A_d, 
\end{equation}
where
$$ 
\begin{gathered}
\chi_\infty (X_\infty )=\case{\exp [-2\pi i(z+\bar z)], &&d<0, \\ 
\exp [-2\pi i(x+x')], &&d>0; }
\\
\chi_p(x)=\exp (2\pi i\{x\}_p), \qquad x\in\QQ_p,
\end{gathered}
$$ 
where $\{x\}_p$ is the fractional part of the $p$-adic number $x$. 

Denote by $r_p$ the rank of the local character $\chi_p (X+\bar X),$ 
$X\in\QQ_p (\sqrt d)$, $p\in S_d$.
Only a finite number of these ranks are different from~0.

Denote by $D$ the discriminant of the field $\QQ (\sqrt d)$:
\begin{equation}
D=\case{d, &&d\equiv 1\ (\mod 4), \\ 4d, &&d\equiv 2,3\ (\mod 4), }
\qq\quad |D|=\prod_{p\in S_d}q_p^{r_p}. 
\end{equation}

The unities $\epsilon$ of the quadratic fields $\QQ (\sqrt d)$ are [4] 
$\epsilon =\pm 1$ and, in addition,
$$ \begin{aligned}
&\epsilon =\pm i \q\hbox{ for } d=-1,  \\
&\epsilon =\pm e^{\pm i{\pi/3}}=\pm{{(1\pm\sqrt{-3})}/ 2} 
	\q\hbox{ for } d=-3, \\
&\epsilon =\pm\Omega^\g,\ \g\in Z,  \q\hbox{ for } d>0,  
\end{aligned}
$$ 
where $\Omega$ is the primary unity of the field $\QQ (\sqrt d)$, 
$N(\Omega )=\Omega\bar\Omega=\pm 1$, $\Omega >0$.

All one-class quadratic fields $\QQ(\sqrt d)$ are well known [4]: there 
exist nine imaginary fields of this type with $d=-1,-2,-3,-7,-11,-19,
-43,-67,-163$ and an infinite number of real fields with 
$d=p$, $p\equiv 3\ (\mod 4)$; $d=2p$, $p\equiv 3\ (\mod 4)$; and 
$d=pp'$, $p,p'\equiv 3\ (\mod 4)$ ($p$~and~$p'$ are prime numbers).

For one-class fields, the group of classes of divisors is trivial; i.e., 
all these divisors belong to the ring of integers (to the maximum order), 
and the prime factor decomposition in this field is unique~[4].

\def\eqnum{\hfill\refstepcounter{equation}$(\theequation)$}

\begin{prp*}[{\mdseries see [4, 18]}]
All prime divisors ${\mathfrak p}\equiv{\mathfrak p}_p$ and 
$\bar{\mathfrak p}\equiv\bar{\mathfrak p}_p$ of one-class quadratic 
fields $\QQ (\sqrt d)$ and their norms $q\equiv q_p$ are classified under 
the following cases\/$:$

\smallskip
{\rm(a)} $p$ divides $D,$ $p \in S_d,$\quad $p=\epsilon{\mathfrak p}^2,$\q 
$q=p;$ \eqnum

\smallskip
{\rm(b)} $p$ does not divide $D,$ $p \in S_d,$\quad $p={\mathfrak p},$\q
$q=p^2;$ \eqnum

\smallskip
$\mathrm{(c')}$ $p$ does not divide $D,$ $p \in P_d,$ $D=4d,$\q 
${\mathfrak p}, \bar{\mathfrak p}=a\pm\sqrt db,$\q $q=p,$\eqnum

\smallskip\noindent
where $a > 0$ and $b > 0$ are integer solutions to the Diophantine equation
\begin{equation}
p=|x^2-dy^2|; 
\end{equation}

$\mathrm{(c'')}$ $p$ does not divide $D,$ $p\in P_d,$ $D=d,$\q 
${\mathfrak p}, \bar{\mathfrak p}=a+{\dfrac b2}\pm\sqrt d{\dfrac b2},$\q 
$q=p,$ \eqnum

\smallskip\noindent
where $a>0$ and $b\neq 0$ are integer solutions to the Diophantine equation
\begin{equation}
p=\left|x^2+xy+{{1-d}\over 4}y^2\right| 
\end{equation}
$($for $d=-7$ and $p=2$, the solution is $a=0,\ b=1)$.  
\end{prp*}

Here, in cases (c$'$) and (c$''$), the divisors ${\mathfrak p}$ and
$\bar{\mathfrak p}$ can be chosen so that the conditions
\begin{equation}
|{\mathfrak p}|_p={1/p}, \qquad |\bar{\mathfrak p}|_p=1 
\end{equation}
are fulfilled.

\begin{rem}
For one-class quadratic fields, the Diophantine equations (2.11) and 
(2.13) are solvable for any prime $p$.
\end{rem}

The following  two theorems give the necessary and sufficient conditions 
for the triviality of the quasicharacter of the idele group of a field on 
the multiplicative group of this field.

\begin{teo}[{\mdseries see [8]}]
The quasicharacter
\begin{equation}
\T(X;\a)={\sgn}^\nu x_\infty\prod_{p=2}^\infty\t_p(x_p)|x_p|_p^{i\a_p}
|X|^\a, \qquad \nu\in F_2,\q\a\in\CC, 
\end{equation}
$(\t_p(p)=1)$ of the group $A^\times$ of ideles 
$X=(x_\infty,x_2,\ldots,x_p,\ldots)$ of the field~$\QQ$ is 
trivial on the group~$\QQ^\times$ if and only if the following conditions 
hold\/$:$
\begin{equation}
\t(-1)=1, \qquad \t(p)=p^{i\a_p},\q p=2,3,\ldots, 
\end{equation}
where $\t$ is a character of the group $A^\times$ of the form
\begin{equation}
\t(X)={\sgn}^\nu x_\infty\prod_{p=2}^\infty\t_p(x_p). 
\end{equation}
\end{teo}

\noindent
{\bf Example}\hspace{3mm}%
of a ramified character of the group $A^\times/\QQ^\times$:
$$ \t(X)=\sgn x_\infty\, \t_2(x_2)\prod_{p=3}^\infty |x_p|^{i\a_p}, 
\qquad \t_2(-1)=-1,\q \t_2(p)=p^{i\a_p},\q p=3,5,\ldots~.$$

\begin{teo}[{\mdseries see [18]}]
The quasicharacter $\T$ of the idele group $A_d^\times$ of 
the one-class quadratic field $\QQ (\sqrt d)$ is trivial on the group 
$\QQ^\times (\sqrt d)$ if and only if the following conditions hold\/$:$
\begin{gather}
 \t (-1)=1\q \hbox { for all admissible } d; \\
 \t (i)=1\q \hbox { for } d=-1; \\
 \t (e^{i{\pi/3}})=1\q \hbox { for } d=-3; \\
 \t (\Omega )=\Omega^{-ia}\q \hbox { for all admissible } d>0; 
 	\\
 p^{i\a_p}=\t ({\mathfrak p})|{\mathfrak p}|^{iah(d)}, \q
 p^{i{\a}'_p}=\t (\bar{\mathfrak p})|\bar{\mathfrak p}|^{iah(d)}, 
 \qquad p\in P_d; \\
 p^{i\a_p}=\t ({\mathfrak p})|{\mathfrak p}|^{iah(d)}, \qquad 
 p \hbox{ divides } D, \quad p\in S_d; \\
 p^{2i\a_p}=\t (p)p^{iah(d)}, \qquad p \hbox{ does not divide } D, 
 \quad p\in S_d; 
\end{gather}
here $\t$ denotes a character of the idele group $A_d^\times,$
\begin{gather} 
\t (X)=\t_\infty (X_\infty)\prod_{p\in P_d\cap R}\t_p(x_p){\t}'_p({x}'_p)
\prod_{p\in S_d\cap R}\t_p(X_p), 
\\
\t_\infty (X_\infty )=\case{ z^\nu(z\bar z)^{-\nu/2}, &&\nu\in Z, \ d<0, \\ 
\sgn^\nu x\cdot\sgn^{\nu'}x', &&\nu, \nu'\in F_2, \ d>0}
\end{gather} 
$(h(d)=1$ for $d>0,$ $h(d)=0$ for $d<0)$.
\end{teo}

\begin{rem}
Equation (2.21) ambiguously determines the real number $a$:
$$ a=-{{\Arg\t ( \Omega)}\over{\ln\Omega}}+{{2\pi}\over{\ln\Omega}}Z.$$ 
Therefore, the real numbers $\a_p$ and ${\a}'_p$ are determined by 
equalities (2.22)--(2.24) ambiguously. However, in what follows, we will 
need only the exponential functions 
$p^{i\a_p}$ and~$p^{i{\a}'_p}$, $p\in P_d$, and~$q^{i\a_p}$, $p\in S_d$,
of these numbers; for $d>0$, these exponential functions are 
determined uniquely.
\end{rem}

\begin{rem}
For the principal quasicharacter $|X|^\a$, $\a\in\CC$, conditions 
(2.18)--(2.24) hold trivially by virtue of the known Artin multiplication 
formula $|I_x|=1$, $x\in\QQ_p^\times (\sqrt d)$.
\end{rem}

\sek{ADELIC FORMULAS FOR GAMMA AND BETA\\ FUNCTIONS OF ONE-CLASS QUADRATIC 
FIELDS\\ AND A FIELD OF RATIONAL NUMBERS}

Taking into account equality (2.18), we can represent the regularized adelic 
formulas of one-class quadratic fields $\QQ(\sqrt d)$ for gamma functions 
of the quasicharacter $\T (X;\a)$ (defined by~(2.1)) of the idele group 
$A_d^\times /\QQ^\times (\sqrt d)$ as [16, 17]
\begin{equation}
\t(-1)=\reg\[\prod_{p=2}^\infty \G(\a+i\a_p;\t_p)
\prod_{p\in P_d\cap F}\G(\a+i{\a}'_p;{\t}'_p)\]
\case{\G_\omega (\a;\nu),&& d<0, \\ \G_\infty (\a+ia;\nu)\G_\infty 
(\a;\nu'), &&d>0,}\q\ 
\end{equation}
where the character $\t(X)$ is defined by (2.25) and
\begin{gather}\nn
\G_\omega (\a;\nu)=\int_{\CC}z^{\nu}(z\bar z)^{-{\nu/2}+\a-1}\exp 
[-2\pi i(z+\bar z)]|dz\wedge d\bar z|\\
=i^{-\nu}2(2\pi)^{-2\a} \G\!\(\a+{\nu\over2}\)\G\!\(\a-{\nu\over2}\)
\sin{\pi\over 2}(2\a-\nu), \qquad \nu\in Z, 
\end{gather}
is the gamma function of quasicharacter $z^{\nu}(z\bar z)^{-{{\nu}/2}+\a}$ 
of the field~$\CC$;
\begin{equation}
\G_\infty (\a;\nu )=\int_{\RR}{\sgn}^{\nu}x|x|^{\a-1}\exp (-2\pi ix)\,dx
=2i^{-\nu}(2\pi)^{-\a}\G(\a)\cos {{\pi}\over 2}(\a-\nu), \qquad 
\nu\in F_2, 
\end{equation}
is the gamma function of quasicharacter $\sgn^{\nu}x |x|^\a$ of the field 
$\RR$ ($\G (\a )$ is the Euler gamma function);\vadjust{\kern-3mm}
{\def\theequation{\ifnum\value{equation}=5\relax\thesection.$4'$%
	\setcounter{equation}{4}\else\thesection.\arabic{equation}\fi}
\begin{gather}
\G (\a;\t)=q^{-{r\over2}}\int_{\KK}\t(x)|x|^{\a-1}\chi(x)\,dx =
\case{q^{(\a-{1/2})r}\G_q(\a), &&p\in F,\\
\kappa(\t)q^{(\a-{1/2})(r+\rho)}, &&p\in R;}
\\ 
\G_q(\a)=G_q(q^\a), \qq \a\neq{{2\pi i}\over{\ln q}}Z, \qquad 
G_q(x)={{1-{x/q}}\over{1-x^{-1}}}, 
\end{gather}
is the reduced gamma function of a local $p$-field with the module $q$;} 
and
$$ \kappa(\t)=q^{\rho/2}\int_{|x|=1}\t(x)\chi (\pi^{-r-\rho}x)dx, \qquad 
|\kappa(\t)|=1.$$ 

By Theorem 2 (see Section 2), equalities (2.18)--(2.24) are valid for  
one-class quadratic fields. Using these equalities, we eliminate the 
numbers $\a_p$ and ${\a}'_p$ from the gamma functions in adelic 
formulas (3.1). As a result, we obtain the adelic formulas
\begin{equation}
\begin{gathered}
\t(-1)\kappa {(N|D|)}^{{1/2}-\a}\\=\reg\prod_{p\in F}\G_q(\a+i\a_p)
\prod_{p\in{P_d\cap F}}\G_p(\a+i{\a}'_p)
\case{\G_\omega(\a;\nu), &&d<0,\\ \G_\infty (\a+ia,\nu)\G_\infty(\a,{\nu}'),
 &&d>0,} 
\end{gathered}
\end{equation}
where the numbers $N$ and $D$ are defined by formulas (2.5) and (2.7),
\begin{equation}
\kappa=\prod_{p=2}^\infty\kappa (\t_p)q_p^{-i\a_p(r_p+\rho_p)}
\prod_{p\in{P_d\cap R}}\kappa ({\t}'_p)p^{-i{\a}'_p{\rho}'_p}, \qq
|\kappa|=1, 
\end{equation}
where the quantities $q_p^{-i\a_p}$ and $p^{-i{\a}'_p}$ are determined by 
equalities (2.22)--(2.24), while $\G_q(\a+i\a_p)$ and $\G_p(\a+i{\a}'_p)$ 
are determined by the formulas
\begin{align}
\G_p (\a+i\a_p )&=G_p [p^{\a}\t({\mathfrak p}){\mathfrak p}^{ih(d)a}], 
\\ 
\G_p(\a+i{\a}'_p)&=G_p [p^\a\t (\bar{\mathfrak p})\bar{\mathfrak p}
^{ih(d)a}], \qq p\in P_d\cap F,\\ 
\G_q (\a+i\a_p )&=G_q [q^\a\t ({\mathfrak p})p^{ih(d)a}], \qquad 
p\in S_d\cap F. 
\end{align}
(In (3.6), we assume that $\kappa(\t_p)=1$ for $p\in F$.)

Suppose that, in addition to the quasicharacter $\T (X;\a)$, another 
quasicharacter $\Pi (Y;\b)$ of the idele group $A_d^\times/\QQ^\times(\sqrt 
d)$ is defined,
\begin{equation}
\Pi (Y;\b)=\pi_\infty (Y_\infty )
\prod_{p\in P_d}
\pi_p (y_p){\pi}'_p({y}'_p)|y_p|_p^{i\b_p}|{y}'_p|_p^{i{\b}'_p}
\prod_{p\in S_d}\pi_p (Y_p)|Y_p\bar Y_p|_p^{i\b_p}|Y|^\b, \qquad 
\b\in\CC, 
\end{equation}
where $b, \b_p,$ and ${\b}'_p$ are real numbers and
\let\z\zeta
\begin{equation}
\pi_\infty (Y_\infty )=\case{\z^{\mu}(\z\bar \z)^{-{\mu}/2},&&\mu\in Z,\ 
d<0, \\ \sgn^\mu y\cdot\sgn^{\mu'}y'|y|^{ib}, &&\mu, \mu'\in F_2, \ d>0.} 
\end{equation}

Let us construct, by formulas (2.25) and (2.26), the character $\pi (Y)$ 
corresponding to the quasicharacter~$\Pi$ and then the character 
$\s (Z)=\bar\t (Z)\bar\pi (Z)$, so that
\begin{equation}
\t\pi\s=1, \qq \s_p=\bar{\t}_p\bar{\pi}_p,\q p=2,3,\ldots, \qquad 
\s'_p=\bar{\t}'_p\bar{\pi}'_p,\q p\in P_d. 
\end{equation}
Denote by $R, R', R''$ and $F, F', F''$ the sets of ramified and 
unramified points of the characters $\t, \pi,$ and~$\s$, respectively.

The beta functions of local fields are defined as the following products of 
three gamma functions:
\begin{equation}
\begin{gathered}
\B_\omega (\a,\nu;\b,\mu;\g,\eta)=\G_\omega (\a;\nu)\G_\omega (\b;\mu)
\G_\omega (\g;\eta),\\
\a+\b+\g=1, \qquad \nu+\mu+\eta =0, \qquad \nu, \mu, \eta\in Z, 
\end{gathered}
\end{equation}
for the quasicharacters $z^\nu (z\bar z)^{-{\nu/2}+\a}$ and  
$\zeta^\mu (\zeta\bar \zeta)^{-{\mu/2}+\b}$, $\nu, \mu \in Z$, of the 
field~$\CC$;
\begin{equation}
\begin{gathered}
\B_\infty (\a,\nu;\b,\mu;\g,\eta)=
\G_\infty (\a;\nu)\G_\infty (\b;\mu)\G_\infty (\g;\eta),\\
\a+\b+\g=1, \qquad \nu+\mu+\eta=0, \qquad \nu,\mu,\eta \in F_2, 
\end{gathered}
\end{equation}
for the quasicharacters $\sgn^{\nu}x|x|^\a$ and $\sgn^{\mu}y|y|^\b$, 
$\nu, \mu \in F_2$, of the field $\RR$; and
\begin{equation}
\B(\a,\t;\b,\pi;\g,\s)=\G(\a;\t)\G(\b;\pi)\G(\g;\s), \qquad 
\a+\b+\g=1, \q \t\pi\s=1, 
\end{equation}
for the local quasicharacters $|x|^{\a}\t(x)$ and $|y|^{\b}\pi(y)$ of the 
$p$-fields.

In particular, for the principal quasicharacters $|x|^\a_p$, $|y|^\b_p$, 
$p\in P_d\cap F$, $|X\bar X|^\a_p$, $|Y\bar Y|^\b_p$, $p\in S_d\cap F$, 
of the $p$-fields with the module $q=q_p$, the beta function is defined as
\begin{equation}
\B_q(\a,\b,\g)=\G_q(\a)\G_q(\b)\G_q(\g). 
\end{equation}

By virtue of (3.1) and (3.13)--(3.15), the  local beta functions defined by 
the quasicharacters $\T (X;\a)$ and $\Pi (Y;\b)$ of the idele group 
$A_d^\times/\QQ^\times (\sqrt d)$ satisfy the following adelic equality 
(for $\a+\b+\g =1$):
\begin{equation}
\begin{gathered}
1=\reg\[\prod_{p=2}^\infty \B(\a+i\a_p,\t_p;\b+i\b_p,\pi_p;\g+i\g_p,\s_p)
{}\right.\\ \left.\times
\prod_{p\in P_d\cap (F\cup F'\cup F'')}\B(\a+i{\a}_p^\sharp,{\t}_p^\sharp;
\b+i{\b}_p^\sharp,{\pi}_p^\sharp;\g+i{\g}_p^\sharp,{\s}_p^\sharp)\]
{}\\  \times
\case{\B_\omega (\a,\nu;\b,\mu;\g,\eta),\q \nu+\mu+\eta=0,\ 
	\nu,\mu,\eta\in Z,\ d<0,\\ \B_\infty(\a+ia,\nu;\b+ib,\mu;\g+ic,\eta)
	\B_\infty(\a,\nu';\b,\mu';\g,\eta'),}\\ 
	\nu+\mu+\eta=\nu'+\mu'+\eta'=0,\qq  
	\nu, \mu, \eta, \nu', \mu', \eta'\in F_2,\\
a+b+c=0,\qq a, b, c\in \RR, \qq d>0;
\end{gathered}
\end{equation}
here, ${\a}_p^\sharp={\a}'_p$ and ${\t}_p^\sharp={\t}'_p$ if $p\in F$, 
${\a}_p^\sharp=\a_p$ and ${\t}_p^\sharp=\t_p$ if $p\in R$, etc.
The real numbers $\a_p, \b_p, \g_p, \a'_p, \b'_p,$ and~$\g'_p$ in~(3.17) 
must satisfy the relations
$$ \a_p+\b_p+\g_p=0, \q p=2,3,\ldots, \qquad 
\a'_p+\b'_p+\g'_p=0, \q p\in P_d.$$ 

Using the adelic formula (3.5), we rewrite (3.17) in expanded form in 
terms of the reduced beta function~$\B_q$; the real numbers 
$\a_p,\b_p,\g_p,{\a}'_p,{\b}'_p,$ and~${\g}'_p$ are eliminated from these 
beta functions by the following relations:\vadjust{\kern-3mm}
\begin{gather}\nn
\B_p(\a+i\a_p,\b+i\b_p,\g+i\g_p)\\
=G_p[p^\a\t ({\mathfrak p}){\mathfrak p}^{ih(d)a}]
G_p[p^\b\pi ({\mathfrak p}){\mathfrak p}^{ih(d)b}]
G_p[p^\g\s ({\mathfrak p}){\mathfrak p}^{ih(d)c}],\qq p\in P_d\cap F, 
\\[2mm]
\B_p(\a+i{\a}'_p,\b+i{\b}'_p,\g+i{\g}'_p)\nn \\
=G_p[p^\a\t (\bar{\mathfrak p})\bar{\mathfrak p}^{ih(d)a}]
G_p [p^\b\pi (\bar{\mathfrak p})\bar{\mathfrak p}^{ih(d)b}]
G_p[p^\g\s (\bar{\mathfrak p})\bar{\mathfrak p}^{ih(d)c}], 
\qq p\in P_d\cap F,
\\[2mm] 
\B_q(\a+i\a_p,\b+i\b_p,\g+i\g_p)\nn \\
=G_q[q^{\a}\t ({\mathfrak p})p^{ih(d)a}]G_q[q^\b\pi ({\mathfrak p})
p^{ih(d)b}]G_q[q^\g\s ({\mathfrak p})p^{ih(d)c}],\qq p\in S_d\cap F. 
\end{gather}

Note that, for $d > 0$, the following equalities hold in (3.17) in 
accordance with the formulas (2.21) and (3.12):
\begin{equation}
\Omega^{ia}=\bar{\t}(\Omega), \qquad 
\Omega^{ib}=\bar{\pi}(\Omega), \qquad 
\Omega^{ic}=\bar{\s}(\Omega)=\t(\Omega)\pi(\Omega); 
\end{equation}
therefore, $a + b + c = 0$ in this case as well.

For the principal quasicharacters $|X|^\a$ and $|Y|^\b$ ($\nu=\mu=\eta=0$,
$\nu'=\mu'=\eta'=0$, $a=b=c=0$, $\a_p=\b_p=\g_p=0$, $\a'_p= \b'_p=\g'_p=0$,
$\kappa=1$, $N(J)=1$, and $R$ is an empty set),
the adelic formula (3.17) for an arbitrary quadratic field $\QQ (\sqrt d)$ 
is rewritten as~[19]
\begin{equation}
\sqrt{|D|}=\reg\[\prod_{p\in P_d}\B_p^2 (\a,\b,\g )\prod_{p\in S_d}
\B_q(\a,\b,\g )\]
\case{\B_\omega (\a,\b,\g ), &&\a+\b+\g=1, \ d<0,\\ 
\B_\infty^2(\a,\b,\g ),&& \a+\b+\g=1, \ d>0,}\q 
\end{equation}
where
{\def\theequation{\ifnum\value{equation}=24\relax\thesection.$23'$%
	\setcounter{equation}{23}\else\thesection.\arabic{equation}\fi}
\begin{align}
\B_\infty (\a,\b,\g)&=\B_\infty (\a,0;\b,0;\g,0),\\ 
\B_\omega (\a,\b,\g)&=\B_\omega (\a,0;\b,0;\g,0). 
\end{align}

In particular, for the Gauss field $\QQ (\sqrt{-1})$, this formula is 
represented as}
\begin{equation}
\B_\omega (\a,\b,\g )\B_2(\a,\b,\g )\reg\[\prod_{p\equiv 1(4)}
\B_p^2(\a,\b,\g )\prod_{p\equiv 3(4)}\B_{p^2}(\a,\b,\g )\]=2,\qq
\a+\b+\g=1. 
\end{equation}

For the field $\QQ$ of rational numbers, we proceed similarly and in a 
more simple manner.

By Theorem 1 (see Section 2), the adelic formula for the gamma functions 
of the field~$\QQ$ that are determined by the quasicharacter $\T(X;\a)$ 
of the form (2.15) is given by [16, 17] (cf.~(3.5))
\begin{equation}
\G_\infty (\a;\nu)\reg\prod_{p\in F}\G_p
(\a+i\a_p)=\t(-1)\prod_{p\in R}\kappa(\t_p)p^{-i\a_p\rho_p}N^{{1/2}-\a}, 
\end{equation}
where
\begin{equation}
\G_p(\a+i\a_p)={{1-p^{\a-1}\t(p)}\over{1-p^{-\a}\bar{\t}(p)}}, \qq
p\in F. 
\end{equation}
(The character $\t(X)$ is defined by (2.17).)

For the quasicharacters $\T(X;\a)$ and $\Pi(Y;\b)$ of the form (2.15), the 
adelic formula for the beta functions of the field $\QQ$ is given by [16, 17]
\begin{equation}
\begin{gathered}
\B_\infty (\a,\nu;\b,\mu;\g,\eta )\reg\prod_{p\in F\cap F'\cap F''}
\B_p(\a+i\a_p,\b+i\b_p,\g+i\g_p)=\kappa(NN'N'')^{{1/2}-\a}, \\
\a+\b+\g=1, \qquad \nu+\mu+\eta=0,\qq \nu, \mu, \eta\in F_2, 
\end{gathered}
\end{equation}
where the characters $\t$ and $\pi$ are constructed by the quasicharacters 
$\T$ and~$\Pi$ by formula~(2.17),  $\t\pi\s=1$;
$\rho_p, {\rho}'_p, {\rho}''_p$ and $N, N', N''$ are the ranks and the 
norms of the leading ideals of the characters $\t, \pi,$ and~$\s$
of the field~$\QQ$, respectively; and
\begin{equation}
\kappa=\prod_{p\in R}\kappa (\t_p)p^{-i\a_p\rho_p}\prod_{p\in F'}\kappa 
(\pi_p)p^{-i\b_p{\rho}'_p}\prod_{p\in F''}\kappa 
(\s_p)p^{-i\g_p{\rho}''_p}, \qq |\kappa|=1. 
\end{equation}
By virtue of (2.16), (3.16), and (3.26),
\begin{equation}
\B_p (\a+\a_p,\b+\b_p,\g+\g_p)
={{1-p^{\a-1}\t(p)}\over{1-p^{-\a}\bar\t(p)}}{{1-p^{\b-1}\pi(p)}
\over{1-p^{-\b}\bar\pi(p)}}{{1-p^{\g-1}\s(p)}\over{1-p^{-\g}\bar\s(p)}}, 
\qquad p\in F, 
\end{equation}
in formula (3.27).

For the principal quasicharacters $|X|^\a$ and $|Y|^\b$ ($\nu=\mu=\eta=0$, 
$\a_p=\b_p=\g_p=0$, and $N=1$) formula (3.27) takes the form~[20]
\begin{equation}
\B_\infty (\a,\b,\g)\reg\prod_{p=2}^\infty \B_p(\a,\b,\g)=1, \qquad 
\a+\b+\g=1. 
\end{equation}

To sum up, we present the following theorems.

\begin{teo}[{\mdseries see [17]}]
Suppose that the quasicharacters $\T$ and $\Pi$ of 
the form\/~$(2.15)$ are given on the idele class group of the field~$\QQ,$
and let $\t, \pi,$ and~$\s,$ $\t\pi\s=1,$ be the characters, of 
the form\/~$(2.17),$ of the idele group of the field~$\QQ$. Then, the adelic 
formulas {\rm(3.25), (3.27)} are valid for local gamma and beta functions.
\end{teo}

\begin{teo}
Suppose that the quasicharacters $\T$ and $\Pi$ of the form\/~$(2.1)$ are 
given on the idele class group of the one-class field $\QQ (\sqrt d),$ and 
let $\t, \pi,$ and~$\s,$ $\t\pi\s=1,$ be the characters, of the 
form\/~$(2.25),$ of the idele group of the field $\QQ (\sqrt d)$. Then, the 
adelic formulas $(3.5)$ and\/ $(3.17)$ are valid for local gamma and beta 
functions.
\end{teo}

\begin{rem}
The adelic formulas (3.1) and (3.17) for the gamma function 
$\G_{A_d}(\a;\ldots)$ and the beta function $\B_{A_d}(\a,\b,\g;\ldots)$ of 
the adele ring $A_d$ of the field $\QQ(\sqrt d)$ imply that 
$\G_{A_d}(\a;\ldots)=\t(-1)$ and $\B_{{A_d}}(\a,\b,\g;\ldots)=1$, 
$\a+\b+\g=1$. Similar expressions are valid for the field~$\QQ$.
\end{rem}

\begin{rem}
In [17, 16], I established the adelic formulas under the condition 
that the ranks of the local characters $\t_p, \pi_p,$ and~$\s_p$, 
$p=2,3,\ldots$, and ${\t}'_p, {\pi}'_p,$ and~${\s}'_p$, $p\in P_d$, 
are identical:
$$ \rho(\t_p)=\rho(\pi_p)=\rho(\s_p),\q p=2,3,\ldots;\qq 
\rho({\t}'_p)=\rho({\pi}'_p)=\rho({\s}'_p),\q p\in P_d.$$ 
In the present work, this requirement is removed at the expense of 
complicating the adelic formulas.
\end{rem}

\noindent
{\bf Example}\hspace{3mm}%
of characters $\t, \pi,$ and $\s=\bar\t\bar\pi$ with identical ranks for 
$p = 5$: $x_5=2^k$, $a\in Z_5^\times$, $|1-a|_5<1$,
$$ \t_5(x_5)=\pi_5(x_5)=\exp(2\pi ik/5),\q k=0,1,2,3; \qquad 
\rho (\t_5)=\rho (\pi_5)=\rho (\bar{\t}_5)=1.$$  

\sek{APPLICATION TO 4-PARTICLE STRING\\ AMPLITUDES}

Suppose that $s, t$, and $u$ are the Mandelstam variables for a 4-particle 
scattering process with the momenta 
$(p_1, p_2, p_3, p_4)$, $p_1+p_2+p_3+p_4=0$, $p_i^2=m_i^2$
in an $n$-dimensional space~$\RR^n$ with the Minkowskian metric 
$p^2=(p^0)^2-(p^1)^2-\ldots-(p^{n-1})^2$, so that 
\begin{equation}
s=(p_1+p_2)^2,\q t=(p_2+p_3)^2,\q u=(p_1+p_3)^2,\qq s+t+u=\sum m_i^2. 
\end{equation}

In the general case of 4-particle scattering processes with the Regge 
trajectory $\a (s)=\a+\a's$, we define the generalized crossing-symmetric 
Veneziano~$V$ and Virasoro~$W$ amplitudes and their 
$p$-adic analogues $V_p$ and $W_p$ by the following formulas 
$(s+t+u=\sum m_i^2)$~[21]:
\begin{align}
V(s,t,u)&=\B_\infty (-\a-\a's,-\a-\a't,-\a-\a'u); 
\\
V_p(s,t,u)&=\B_q (-\a-\a's,-\a-\a't,-\a-\a'u); 
\\
W(s,t,u)&=\B_\omega (-{\a/2}-{\a'/2}s,-{\a/2}-{\a'/2}t,-{\a/2}-{\a'/2}u); 
\\ 
W_p(s,t,u)&=\B_q (-{\a/2}-{\a'/2}s,-{\a/2}-{\a'/2}t,-{\a/2}-{\a'/2}u). 
\end{align}
The slope $\a'$ and the intercept $\a$ of the trajectory $\a(s)=\a+\a's$
must satisfy the following condition:
\begin{equation}
3\a+\a'\sum m_i^2=\case{-1 & for amplitude & V, \\ -2 & for amplitude &W.}
\end{equation}
By virtue of (3.21), the amplitudes (4.2)--(4.5) satisfy the following adelic 
relations (provided that $s+t+u=\sum m_i^2$):
\begin{align}
V^2(s,t,u)\reg\[\prod_{p\in P_d}V_p^2(s,t,u)\prod_{p\in S_d}V_p(s,t,u)\]
&=\sqrt{|D|}, \qquad d>0,\\ 
W(s,t,u)\reg\[\prod_{p\in P_d}W_p^2 (s,t,u)\prod_{p\in S_d}W_p (s,t,u)\]
&=\sqrt{|D|}, \qquad d<0. 
\end{align}
Formulas (4.7), (4.8) are well known [13, 19]. They provide a decomposition 
of generalized Veneziano~$V\ (d>0)$ and Virasoro~$W\ (d<0)$ amplitudes 
into the infinite product of the inverses $V_p^{-1}$ and~$W_p^{-1}$ of the 
$p$-adic amplitudes $V_p$ and~$W_p$, respectively.

Let us present certain important particular cases of the generalized 
amplitudes introduced. These are, first of all, 4-tachyon 
crossing-symmetric string amplitudes for tree orientable 
diagrams\kern0pt---the 
Veneziano and Virasoro amplitudes [11--13, 21--25].

\subsection*{The Veneziano amplitude for open strings in $\RR^{26}$.}

There exist only one tree orientable diagram that is conformally  
equivalent to a unit circle with four punctured points on its boundary. 
This diagrams corresponds to the crossing-symmetric Veneziano amplitude 
(for $\a=1$, $\a'={1/2}$, $m_i^2=-2$, see formula (4.2))
\begin{equation}
V(s,t,u)=\B_{\infty}(-1-{s/2},-1-{t/2},-1-{u/2}), \qquad s+t+u=-8. 
\end{equation}
The amplitudes $V_p$ are determined similarly by formula (4.3).

\subsection*{The Virasoro amplitude for closed strings in $\RR^{26}$.}

There exists only one tree orientable diagram that is conformally  
equivalent to a unit sphere with four punctured points. This diagrams 
corresponds to the crossing-symmetric Virasoro amplitude (for $\a=2$, 
$\a'={1/4}$, $m_i^2=-8$, see formula~(4.4))
\begin{equation}
W(s,t,u)=\B_\omega (-1-{s/8},-1-{t/8},-1-{u/8}), \qquad s+t+u=-32. 
\end{equation}
The amplitudes $W_p$ are determined similarly by formula (4.5).

The Veneziano and Virasoro amplitudes are generalized to the case of 
ramified quasicharacters by the formulas (under conditions (4.1) and (4.6))
\begin{gather}
\begin{gathered}
V_{\nu\mu\eta}(s,t,u)=\B_\infty (-\a-\a's,\nu;-\a-\a't,\mu;-\a-\a'u,\eta),\\
\nu+\mu+\eta =0,\qq \nu,\mu,\eta\in F_2, \qquad d>0; 
\end{gathered}\\ \begin{gathered}
W_{\nu\mu\eta}(s,t,u)=\B_{\omega}(-\a-\a's,\nu;-\a-\a't,\mu;-\a-\a'u,\eta),\\ 
\nu+\mu+\eta =0, \qq \nu,\mu,\eta\in Z, \qquad d<0. 
\end{gathered}
\end{gather}
These amplitudes are crossing symmetric with respect to the permutations 
of the variables $(s,\nu )$, $(t,\mu )$, and $(u,\eta )$ and satisfy the 
adelic formulas of Section~3 (for one-class quadratic fields and a field of 
rational numbers).

\subsection*{The massless 4-particle amplitude of the 
Ramond--Neveu--Schwarz superstring in~$\RR^{10}$.}

This amplitude is proportional to [11, 21]
\begin{equation}
A_\infty (s,t,u)=\G_\infty (-s/2;1)\G_\infty (-t/2;1)\G_\infty (-u/2;1), 
\qquad s+t+u=0. 
\end{equation}
The corresponding simple adelic formula for 
$\nu\,{=}\,\mu\,{=}\,\eta\,{=}\,1$ and
$\rho_p\,{=}\,\rho(\t_p)\,{=}\,\rho(\pi_p)\,{=}\,\rho(\s_p)$, 
$p\in R=R'=R''$,
is rewritten as (see~(3.25))
\begin{equation}
A_\infty (s,t,u)\reg\prod_{p\in F}\G_p(-s/2+i\a_p)\G_p(-t/2+i\b_p)\G_p(-u/2+i\g_p)
=\kappa N\sqrt N, \qquad s+t+u=0, 
\end{equation}
where $\kappa$ is defined by (3.28); the functions 
$\G_p(-s/2+i\a_p),\ldots$ are calculated by (3.26), and the quantities 
$p^{-i\a_p},\ldots$, by formulas of the type (2.16).

Another adelic formula for the amplitude 
$A_\infty (s,t,u)$ was proposed in~[26].

\subsection*{Massless amplitudes for four charged particles of a heterotic 
superstring in~$\RR^{10}$.}

There are four basic types of such amplitudes; namely [21],
\begin{equation}
A_\infty^{(k)} (s,t,u)=\B_\infty (-1-{s/8}-{S/2},-1-{t/8}-{T/2},
-1-{u/8}-{U/2}), 
\end{equation}
where $s+t+u=0$ and $S+T+U=-8$; here, $k=1$ corresponds to the 
set of indices $(S=-8,\ T=0$, and $U=0);$ $k=2$, to $(S=-6,\ T=-2,$ 
and $U=0);$ $k=3$, to $(S=-4,\ T=-4,$ and $U=0);$ and $k=4$, to $(S=-4,\ 
T=-2,\ U=-2)$. The other amplitudes are obtained by the 
permutation of indices $S, T$, and~$U$, $S + T + U = 0$ (that assume the 
values $0, -2, -4, -6,$ and~$ -8$).

The amplitudes $A_\infty^{(k)} (s,t,u)$, $k=1,2,3,4$, are the beta 
functions of the field~$\RR$ (which are not crossing symmetric). 
Therefore, they satisfy any adelic formula from Section~3.

Another type of adelic formulas for these amplitudes is obtained if we 
represent them as the products of the three gamma functions
$$ 
\begin{gathered}
\G_\infty (-{s/8};\nu),\q \G_\infty (-{t/8};\mu), \q
\G_\infty (-{u/8};\eta),\\
s+t+u=0,\qq \nu+\mu+\eta=1,\qq \nu, \mu, \eta\in F_2,
\end{gathered}
$$  
and apply the adelic formula (3.25) to each.

For the calculations, we first apply formulas (3.3):
\begin{align}
S&=-8,\qq \G_\infty (-1-{s/8}+4)=-(16\pi)^{-3}i(16-s)(8-s)s\G_\infty 
(-{s/8};1), 
\\
S&=-6,\qq \G_\infty (-1-{s/8}+3)=(16\pi)^{-2}(8-s)s\G_\infty (-{s/8};0), 
\\
S&=-4,\qq \G_\infty (-1-{s/8}+2)=(16\pi)^{-1}is\G_\infty (-{s/8};1), 
\\
S&=-2,\qq \G_\infty (-1-{s/8}+1)=\G_\infty (-{s/8};0), 
\\
S&=0,\9-\qq \G_\infty (-1-{s/8}+0)={{16\pi i}\over{8+s}}\G_\infty 
(-{s/8};1), 
\end{align}
and then formulas (3.23). As a result, we obtain
\begin{align}
A_\infty^{(1)}(s,t,u)&={{-i}\over{16\pi}}{{(16-s)(8-s)s}\over{(8+t)(8+u)}}\G_\infty (-{s/8};1)\G_\infty (-{t/8};1)\G_\infty (-{u/8};1), 
\\
A_\infty^{(2)}(s,t,u)&={i\over{16\pi}}{{(8-s)s}\over {8+u}}\G_\infty 
(-{s/8};0)\G_\infty (-{t/8};0)\G_\infty (-{u/8};1), 
\\
A_\infty^{(3)}(s,t,u)&={{-i}\over{16\pi}}{{st}\over {8+u}}\G_\infty 
(-{s/8};1)\G_\infty (-{t/8};1)\G_\infty (-{u/8};1), 
 \\
A_\infty^{(4)}(s,t,u)&={{is}\over{16\pi}}\G_\infty (-{s/8};1)\G_\infty 
(-{t/8};0)\G_\infty (-{u/8};0).
\end{align}

Formulas (4.21)--(4.24) and (4.13) show that the amplitudes 
$A_\infty^{(1)}(s,t,u)$ and $A_\infty^{(3)}(s,t,u)$
are proportional to the massless superstring amplitude 
$A_\infty({s/4},{t/4},{u/4}).$

Thus,  we established a relation between the scattering amplitudes in the 
theory of a heterotic string and the Ramond--Neveu--Schwarz superstring.

\begin{rem}
A function of the form (see (4.13), (4.21)--(4.24))
$$ 
\begin{gathered}
\B'_\infty (\a,\nu;\b,\mu;\g,\eta)
=\G_\infty (\a;\nu)\G_\infty (\b;\mu)\G_\infty (\g;\eta),\\ 
\a+\b+\g=0, \qq\nu+\mu+\eta=1, \qq\nu, \mu, \eta\in F_2,
\end{gathered}
$$ 
is not a beta function of the type $\B_\infty$ that was considered in 
Section~3. It as if complements~$\B_\infty$. We denote this function by 
$\B'_\infty$ and call a {\it primed\/} beta function. As we have seen, 
it describes the superstring amplitudes. The integral representation of 
this function is given by
$$ 
\B'_\infty (\a,\nu;\b,\mu;\g,\eta)=-\pi i\bigl ({\sgn}^\nu x 
|x|^{\a-1}\bigr )\ast\bigl ({\sgn}^\mu x |x|^{\b-1}\bigr )\ast\bigl 
({\sgn}^\eta x |x|^{\g-1}\bigr )\Big|_{x=1},$$ 
where $*$ denotes convolution. We will dwell on this point in a different 
time and different place.
\end{rem}

\subsection*{The amplitudes $V_{\nu\mu\eta}$.} 

In total, there exist four solutions to the equation 
$\nu +\mu +\eta =0$, $\nu,\mu,\eta \in F_2$: 000, 110, 101, and~011. 
Therefore, by~(4.11), there exist four different 
amplitudes~$V_{\nu\mu\eta}$. One of them, $V_{000}=V$, is the Veneziano 
amplitude, and the rest three quantities $V_{110}, V_{101},$ and~$V_{011}$
form a 3-vector in which each component is expressed in terms of another, 
for example,
$$ 
V_{101}(s,t,u)=V_{110}(s,u,t)=V_{011}(t,s,u);
$$ 
and all of them are expressed in terms of the Veneziano amplitude $V$, for 
example
\begin{equation}
V_{110}(s,t,u)={{1+\a+\a't}\over{\a+\a's}}V\!\(s-{1\over{\a'}},t
+{1\over{\a'}},u\), 
\end{equation}
by virtue of the easily verifiable relation
$$ 
\B_\infty(\a,1;\b,-1;\g,0)=
{{\b-1}\over\a}\B_\infty (\a+1,\b-1,\g),\qq \a+\b+\g=1.
$$ 

\subsection*{The amplitudes $W_{\nu\mu\eta}$.}

There exist an infinite number of amplitudes~$W_{\nu\mu\eta}$ (4.12), 
exactly as many as there are integer solutions to the equation 
$\nu +\mu +\eta =0$, $\nu,\mu,\eta\in Z$: 
$W_{000}=W$ is the Virasoro amplitude; 
the quantities $W_{\nu\nu\mu}, W_{\nu\mu\nu}$, and  
$W_{\eta\nu\nu},$ $2\nu +\eta =0$, $\nu\neq\eta$, form a 3-vector; and the 
quantities $W_{\nu\mu\eta},W_{\nu\eta\mu}, \ldots ,W_{\eta\mu\nu}$, 
$\nu +\mu +\eta =0$, $\nu\neq\mu\neq\eta\neq\nu$,
form a 6-vector. In each of these groups of vectors, each component is 
expressed in terms of another, for example,
$$ W_{\mu\nu\eta}(s,t,u)=W_{\nu\mu\eta}(t,s,u)=\ldots =W_{\eta\mu\nu}(u,s,t).$$ 
The physical sense of these amplitudes is note yet clear.

\smallskip
Everything that was said in this section concerning one-class quadratic 
fields may also apply to any one-class fields of algebraic numbers. The 
appropriate adelic formulas were obtained in [14--17].

\section*{ACKNOWLEDGMENTS}
I am grateful to I.V.~Volovich for fruitful discussions.

This work was supported by the Russian Foundation for Basic Research (project 
no.~96-01-01008) and the Program ``Leading Scientific Schools of the 
Russian Federation" (project no.~96-15-96131).

\end{document}